\documentclass[letterpaper]{article} 
\usepackage{aaai25}
\usepackage{times}  
\usepackage{helvet}  
\usepackage{courier}  
\usepackage[hyphens]{url}  
\usepackage{graphicx} 
\usepackage{natbib}  
\usepackage{caption} 
\usepackage{algorithm}
\usepackage{algorithmic}
\usepackage{newfloat}
\usepackage{listings}
\DeclareCaptionStyle{ruled}{labelfont=\mathrm{norm}alfont,labelsep=colon,strut=off} 
\floatstyle{ruled}
\newfloat{listing}{tb}{lst}{}
\floatname{listing}{Listing}
\usepackage{booktabs}
\usepackage{multirow}
\usepackage{amssymb}
\usepackage{utfsym}
\usepackage{url}
\usepackage{tablefootnote}
\usepackage{longtable}
\usepackage{amsmath} 

\newcommand{\NumOfTask}{9 } 
\newcommand{\spm}[1]{\textcolor{gray}{\scriptsize{$\pm #1$}}} 

\title{MuQ: Self-Supervised Music Representation Learning \\ with Mel Residual Vector Quantization}
\author {
    Haina Zhu\textsuperscript{\rm 1}\footnote{Work performed during an internship at Tencent AI Lab.},
    Yizhi Zhou\textsuperscript{\rm 2}\footnotemark[1],
    Hangting Chen\textsuperscript{\rm 3}\footnote{Corresponding Authors.},
    Jianwei Yu\textsuperscript{\rm 3}\footnotemark[2],  \\
    Ziyang Ma\textsuperscript{\rm 1}, 
    Rongzhi Gu\textsuperscript{\rm 3}, 
    Yi Luo\textsuperscript{\rm 3}, 
    Wei Tan\textsuperscript{\rm 3}, 
    Xie Chen\textsuperscript{\rm 1}\footnotemark[2]
}
\affiliations {
    \textsuperscript{\rm 1}X-LANCE Lab, Shanghai Jiao Tong University\\
    \textsuperscript{\rm 2}National Key Laboratory for Novel Software Technology, Nanjing University\\
    \textsuperscript{\rm 3}Tencent AI Lab\\
    erichtchen@tencent.com, tomasyu@foxmail.com, chenxie95@sjtu.edu.cn \\

    \begin{links}
    \begin{center}
            \link{\textbf{Code}}{https://github.com/tencent-ailab/MuQ}
            \link{\textbf{Checkpoints}}{https://huggingface.co/OpenMuQ}
    \end{center}
    \end{links}
    
}

\usepackage{bibentry}

\begin{document}

\maketitle

\hyphenation{MusicFM MERT MuQ MuQ-MuLan Projection random-projection re-contruct-ion  Residual tokenizer Quantizer music token tokens code-book code-books MARBLE}

\begin{abstract}

Recent years have witnessed the success of foundation models pre-trained with self-supervised learning (SSL) in various music informatics understanding tasks, including music tagging, instrument classification, key detection, and more. In this paper, we propose a self-supervised music representation learning model for music understanding. Distinguished from previous studies adopting random projection or existing neural codec, the proposed model, named MuQ, is trained to predict tokens generated by Mel Residual Vector Quantization (Mel-RVQ). Our Mel-RVQ utilizes residual linear projection structure for Mel spectrum quantization to enhance the stability and efficiency of target extraction and lead to better performance. Experiments in a large variety of downstream tasks demonstrate that MuQ outperforms previous self-supervised music representation models with only 0.9K hours of open-source pre-training data. Scaling up the data to over 160K hours and adopting iterative training consistently improve the model performance. To further validate the strength of our model, we present MuQ-MuLan, a joint music-text embedding model based on contrastive learning, which achieves state-of-the-art performance in the zero-shot music tagging task on the MagnaTagATune dataset. Code and checkpoints are open source in \url{https://github.com/tencent-ailab/MuQ}.

\end{abstract}

%

\section{Introduction}

\begin{figure}
    \centering
    \includegraphics[width=1.0\linewidth]{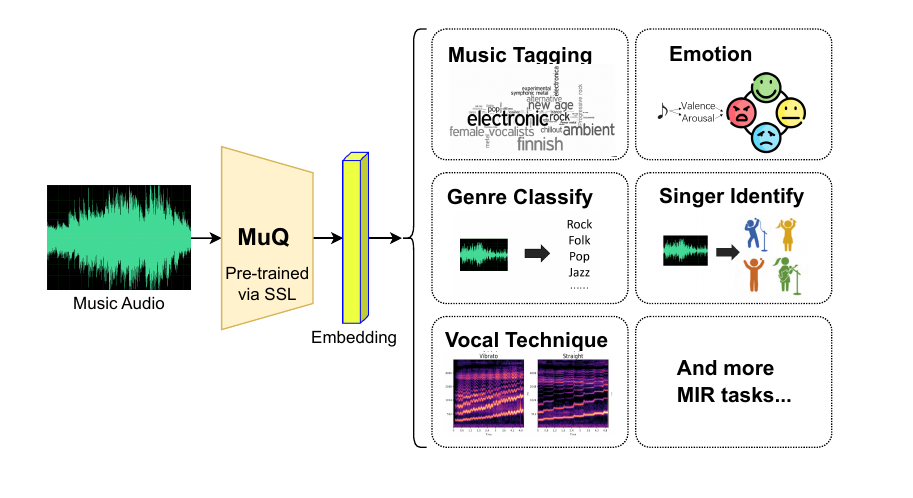}
    \caption{Illustration of how the SSL pre-trained model works in music information
retrieval (MIR) tasks.}
    \label{fig:task_illu}
\end{figure}

Self-supervised learning (SSL) has been introduced into speech and audio signal processing as a technique for learning latent semantic relationships from unlabeled raw data. Recently, several works \cite{li2023mert, won2023musicfm} apply SSL to music informatics understanding, and a number of pre-trained foundation models (PFMs) with generalized representation capabilities are built. Learning from unsupervised music data, these foundation models can achieve stunning performance in music understanding tasks such as genre classification, emotion prediction, and key detection \cite{yuan2023marble}, and further provide semantic representations for more downstream tasks like music generation \cite{agostinelli2023musiclm} and music captioning \cite{deng2024musilingo}, as illustrated in Figure \ref{fig:task_illu}. 

A key challenge in music understanding and representation is that music is an extremely specific modality. Unlike speech or environmental sounds, music not only focuses on semantic information, but also emphasizes acoustic information, such as melody, chords, and tonality. As a result, previous semantics-oriented SSL methods have struggled to perform well on music tasks\cite{spijkervet2021clmr, li2022mapmusic2vec, yuan2023marble}, as they fail to simultaneously capture both semantic and acoustic information.

Recently, several studies have sought to develop a universal music representation that integrates both semantic and acoustic aspects. Among these efforts, two remarkable models are MERT \cite{li2023mert} and MusicFM \cite{won2023musicfm}. MERT employs a BERT-style masked language modeling (MLM) proxy task to predict discrete tokens from the masked audio parts, with an Encodec \cite{defossez2022encodec} model as the tokenizer and uses an auxiliary Constant Q-Transform (CQT) target to enhance the modeling of acoustic information. MusicFM, on the other hand, directly utilizes a random projection quantizer derived from BEST-RQ \cite{chiu2022bestrq}, and this tokenization approach provides a general target for learning music representation, without the need for additional CQT loss to capture acoustic modeling.

As discussed in \cite{chiu2022bestrq} and \cite{li2023mert}, the target extractor (i.e., tokenizer) plays an important role in SSL, as models are trained to predict the tokenized pseudo-labels. BEST-RQ features a lightweight approach that allows for fast extraction of discrete targets. However, its performance is highly dependent on the initialization of the random projection layer, often requiring multiple attempts or a specific random seed to achieve optimal results. In contrast, the Encodec target \cite{defossez2022encodec} used in MERT produces a series of residual targets, with the multi-target strategy shown beneficial to musical SSL \cite{li2023mert}. As a neural codec trained on audio data, Encodec produces more stable labels compared to its random counterparts. However, using Encodec as tokenization is computational heavy and consumes a lot of GPU memory when applying online extraction, which can reduce the training efficiency. Also, it needs to be coupled with additional CQT reconstruction loss to perform well in acoustic representation.


To address the initialization dependency of the random projection quantizer in BEST-RQ and the inefficiency stemming from the heavy computation cost of Encodec in MERT, we introduce a model called MuQ, which learns \textbf{Mu}sic representations from Mel \textbf{Q}uantization targets.
MuQ leverages a Mel-RVQ as the tokenizer to generate targets. The proposed Mel-RVQ is pre-trained on music data and employs a linear RVQ to directly quantize audio Mel spectrograms. Compared to the random-projection quantizer in BEST-RQ, the pre-trained Mel-RVQ produces more stable targets for SSL training and eliminates the model's dependence on initialization. Additionally, compared to Encodec, the lightweight single-layer Mel-RVQ architecture offers greater extraction efficiency. 

To further demonstrate the capabilities of the proposed MuQ model, we explore its application in another crucial area of music understanding: aligning music and text representations. For example, MuLan \cite{huang2022mulan} employs contrastive learning to train both a music encoder and a text encoder, producing semantically consistent embeddings for both modalities and achieving coherent alignment between music and text. Recognizing the role that self-supervised learning models can play in providing effective initialization for downstream tasks, we leverage our MuQ to construct a joint music-text embedding model, named MuQ-MuLan.

Our main contributions
are listed as follows:

\begin{itemize}

\item We introduce a novel music SSL model MuQ, which demonstrates state-of-the-art performance across a wide range of downstream music understanding tasks over previous MERT and MusicFM models.

\item We propose the Mel Residual Vector Quantization (Mel-RVQ), which directly quantizes the Mel spectrum using a single linear layer RVQ, improving both training stability and efficiency.

\item We further develop the MuQ-MuLan model, trained with contrastive learning on MuQ. MuQ-MuLan excels in aligning and jointly encoding music and text modalities, compared with the original MuLan model.

\end{itemize}

Our experiments demonstrate that the Mel-RVQ significantly enhances SSL performance across a variety of music downstream tasks. Notably, MuQ outperforms previous state-of-the-art (SOTA) SSL models MERT and MusicFM, even when trained on just 0.9K hours of data, which is 100x less than what comparable models require \cite{li2023mert, won2023musicfm}.  Additionally, our results show that MuQ-MuLan achieves a ROC-AUC score of 79.3 on the MagnaTagATune zero-shot music tagging task, surpassing the previous SOTA result\cite{huang2022mulan}. 

\section{Related Work}
\paragraph{\textbf{Self-supervised learning for speech and audio}} 
Self-supervised learning (SSL)  \cite{mohamed2022self} has been introduced to the field of speech and audio signal processing as a way to learn semantic representations from unlabeled audio data, and it is now widely used in tasks such as automatic speech recognition \cite{baevski2023d2v2} and audio event classification \cite{chen2024eat}.

SSL often relies on well-designed targets. HuBERT \cite{hsu2021hubert}, for instance, uses K-means clustering labels to guide self-supervised learning in Masked Language Model (MLM) style, with these labels being extracted offline. BEST-RQ \cite{chiu2022bestrq} learns directly from targets generated by a randomly initialized projection quantizer, which can be extracted online. MT4SSL \cite{ma2022mt4ssl} integrates both online and offline objectives, applying the concept of multi-target self-supervised learning, where a distinct linear prediction head is used for each type of target.

Another important topic in SSL is iterative refinement training. Most existing iterative training practices are based on clustering. For example, HuBERT applies K-means clustering on the features of an existing model to train the next iteration of models. Similarly, Seed-ASR \cite{bai2024seedasr} first trains an SSL model on BEST-RQ and then iteratively trains on its features using K-means clustering. 

In this paper, we take an alternative approach by iteratively training directly through the quantizer itself, without the need to apply K-means clustering.

\paragraph{\textbf{Music processing and understanding }} 

Music information retrieval (MIR) contains a large set of tasks to evaluate deep learning models' understanding of music from audio signals or symbolized notation. For example, automatic music tagging task require models to do multi-label classification for tags like genre, instrument, and mood, given a music track \cite{won2020tagging}. Other MIR tasks include key detection,  pitch detection, emotion analysis, etc \cite{yuan2023marble}.

In recent years, SSL has been introduced into music understanding for building universal representational models capable of handling various MIR tasks. Some of these efforts include CLMR \cite{spijkervet2021clmr} based on contrastive learning; MERT \cite{li2023mert} based on MLM proxy task; and more recently, MusicFM \cite{won2023musicfm}. These SSL-based models yield impressive performance across multiple MIR tasks, showing a generalized understanding of music.

\paragraph{\textbf{Music-text joint embedding model}} 
Bridging the gap between music and text modalities, music-text representation models learn joint embeddings from music-text pairwise data. Most of these models use a two-tower structure, consisting of both a music encoder and a text encoder, and are trained by contrastive learning loss. 

LAION-CLAP \cite{laionclap2023} is a powerful open-source music-text encoder, with various model structures and versions trained on extensive data. MuLan \cite{huang2022mulan} is also an exceptional music-text embedding model, but it is not open source for either the model or the data.

 The MuQ-MuLan model introduced in this paper follows the ideas of former work but replaces the music encoder with our proposed MuQ model.

\section{Method}

\begin{figure}
    \centering
    \includegraphics[width=0.8\linewidth]{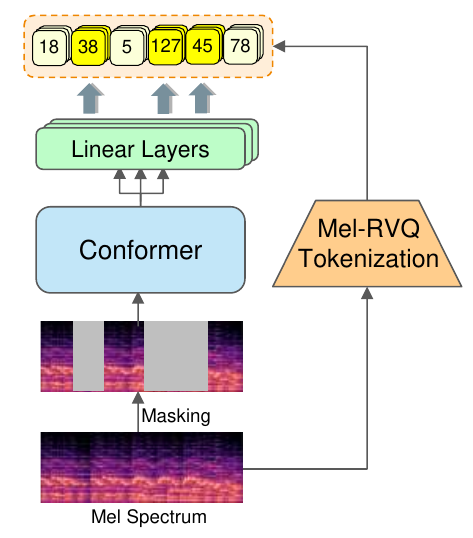}
    \caption{An overview of our proposed MuQ's framework.}
    \label{fig:structure}
\end{figure}
MuQ employs a self-supervised learning approach based on masked language modeling (MLM) and tokenized targets. The overall framework of MuQ is presented in Figure \ref{fig:structure}.

\subsection{Self-Supervised Framework of MuQ}

MuQ directly takes the Mel spectrum of the music audio signal as input.  The Mel spectrum is partially masked as random noise with a masking probability $p$, and then fed into multiple layers of Conformer \cite{gulati2020conformer} for context learning. The Conformer output is passed through linear layers, which serve as prediction heads. Finally, we calculate the cross-entropy loss between the target and predicted labels as the optimization objective.

The target labels are tokens extracted from the Mel spectrum by the Mel-RVQ, a quantization tokenizer proposed in this paper. Since Mel-RVQ produces $N$ tokens for each time step of the Mel spectrum, the MuQ model incorporates $N$ distinct linear layers to predict the different target tokens. In Figure \ref{fig:structure}, these multiple linear layers are illustrated as stacked blocks following the Conformer.

\subsection{Mel Residual Vector Quantization (Mel-RVQ)}

\begin{figure*}[htp]
    \centering
    \includegraphics[width=\textwidth]{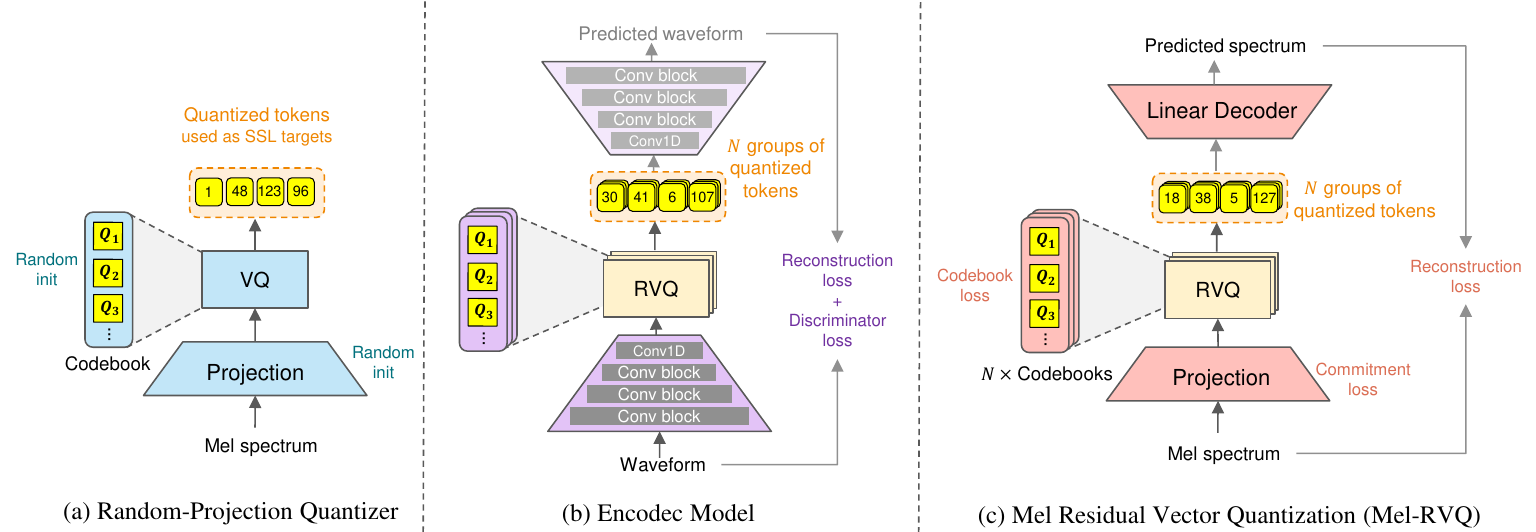}
    \caption{Comparison between  (a) random-projection quantizer, (b)Encodec, and (c) proposed Mel Residual Vector Quantization (Mel-RVQ). Mel-RVQ draws on the lightweight structure of the random-projection quantizer, i.e., it uses only a single linear projection as the encoder, with the Mel spectrum as input. Also, Mel-RVQ borrows the proven effective residual structure from Encodec, but Mel-RVQ further benefits from a much simpler structure than the multiple convolutional layers in Encodec.
    }
    \label{fig:projection}
\end{figure*}

As shown in Figure \ref{fig:projection}, we depict the proposed Mel Residual Vector Quantization (Mel-RVQ) compared with the random-projection quantizer in BEST-RQ and the Encodec target used by MERT.

The proposed Mel-RVQ directly takes the Mel spectrum as input and then quantizes the Mel spectrum using residual vector quantization (RVQ). The encoder of Mel-RVQ is designed as a simple single-layer linear projection, and the decoder is also a single linear layer. 

Mel-RVQ needs to be trained on music data in advance before it can be used to generate quantization targets in SSL training. 

\subsubsection{Training of Mel-RVQ}

During the training of Mel-RVQ, the final loss function can be decomposed into three terms, as shown in Figure \ref{fig:projection}(c) :

\begin{itemize}
    \item \textbf{Codebook loss} trains the RVQ component to improve the fit of the codebook embedding $Q_\tau$ to the dimensionality-reduced features $z$. Formally,
        \begin{equation}
           loss_{\mathrm{code}} = \sum_{ \substack{x \in B, \\ z = M_\mathrm{P} x} } { \| \mathrm{norm}(Q_\tau) - \mathrm{norm}(\mathrm{sg}(z)) \| ^2}
        \end{equation}
    where $B$ refers to a mini-batch of data, $x$ denotes Mel spectrum input and $\mathrm{sg}$ denotes stop-gradient. $M_\mathrm{P}$ is the projection matrix and token $\tau$ is the closest to the projected vector $z$ in the $l_2$-normalized embedding space.
    \item \textbf{Commitment loss} trains the projection (i.e. encoder) for more optimal dimension reduction and fitting to embeddings. Expressed as
        \begin{equation}
           loss_{\mathrm{comm}} = \sum_{ \substack{x \in B, \\ z = M_\mathrm{P} x} }{ \| \mathrm{norm}(z) - \mathrm{norm}(\mathrm{sg}(Q_\tau)) \| ^2}
        \end{equation}
    
    \item \textbf{Reconstruction loss} trains the linear decoder (denoted as $M_\mathrm{D}$) and the codebook to restore the original feature $x$. The formula is
        
        \begin{equation}
           loss_{\mathrm{recon}} = \sum_{x \in B} { \| M_\mathrm{D} Q_\tau - x \| ^2}
        \end{equation}

\end{itemize}

The final loss used to train the Mel-RVQ is the weighted sum of the above three losses:
    \begin{equation}
       loss = \alpha \cdot loss_{\mathrm{code}} + \beta \cdot loss_{\mathrm{comm}} + loss_{\mathrm{recon}}
    \label{eq:final_loss}
    \end{equation}
where $\alpha$ and $\beta$ are weighting factors, set as $\alpha=1, \beta=0.25$.

It is emphasized that both \(M_\mathrm{D}\) and \(M_\mathrm{P}\) used in Eq.(\ref{eq:final_loss}) are simple linear layers, which distinguishes them from those in neural codecs like Encodec \cite{defossez2022encodec}. 

\subsubsection{Residual modeling of Mel-RVQ}

Applying the residual modeling method to Mel-RVQ means that it yields multiple tokens for each time step of the audio. Assume there are $N$ codebooks in total, denoted as $\{Q_i^{(n)}\}_{i=1}^K$ for $n \in 1\ldots N$. In this way, we have
\begin{equation}
\begin{aligned}
   z^{(n)} &= M_\mathrm{P}^{(n)} r^{(n-1)} \\
   \tau^{(n)} &= {\arg\min_{i} \| \mathrm{norm}(z^{(n)}) - \mathrm{norm}(Q^{(n)}_i) \|} \\
   r^{(n)} &= r^{(n-1)} - M_\mathrm{D}^{(n)} Q^{(n)}_{\tau^{(n)}} \\
\end{aligned}
\end{equation}
where  $n \in 1\ldots N$; $r^{(n)}$ is the residual signal passed to next step in quantizer and $r^{(1)} = x$. $M_\mathrm{P}^{(n)}$ and $M_\mathrm{D}^{(n)}$ denote the residual-projection matrices for the encoder and decoder, which contain multiple steps of projection. The superscript $(n)$ denotes the components or features corresponding to the quantizer at step $n$. 

\subsubsection{Iterative refinement with RVQ}

Iterative refinement is introduced in HuBERT \cite{hsu2021hubert} as a method to improve the performance of SSL models, where clustering techniques like K-means are employed to produce new labels for the next iteration of training.

We suggest that residual vector quantization (RVQ) itself can serve as an alternative to the clustering in iterative enhancement.  That is, we directly train a 
Mel-RVQ\textsubscript{iter} on the latent representations of an already trained MuQ model, to re-drive the training of MuQ at next iteration.

Specifically, the iterative training on MuQ has two stages:

\begin{itemize}
    \item In the initial stage, a Mel-RVQ is trained on the Mel spectrum feature, and then the first version of MuQ is trained on the tokens produced by Mel-RVQ.
    \item In the iterative stage, the first version of MuQ is used to extract representation from the audio,
    and the $l$ th layer latent is used to train Mel-RVQ\textsubscript{iter}, which will be used in training the second version of MuQ, namely MuQ\textsubscript{iter}.
\end{itemize}

\subsection{Music-Text Contrastive Learning}

To verify the effectiveness of MuQ in more sophisticated downstream tasks, we trained a music-text joint embedding model, MuQ-MuLan. Similar to MuLan \cite{huang2022mulan}, MuQ-MuLan employs a two-tower multimodal architecture and is trained on a large amount of (music, text) paired data. 

\begin{figure}[h]
    \centering
    \includegraphics[width=0.9\linewidth]{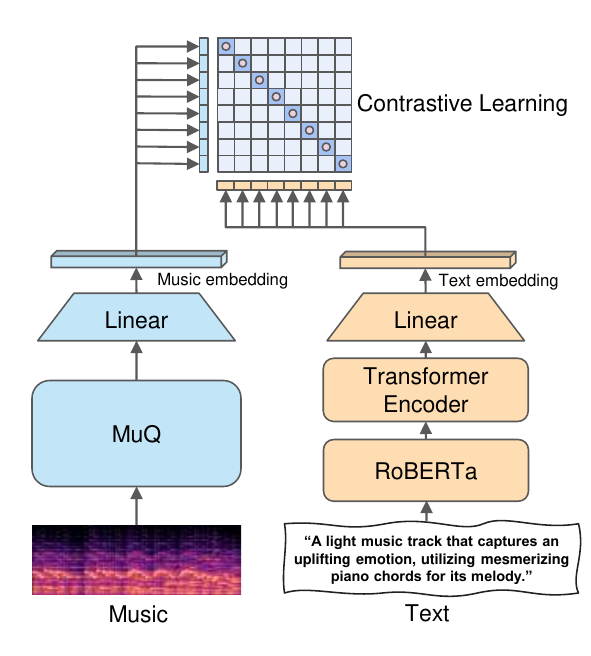}
    \caption{Overall architecture of MuQ-MuLan, a music-text joint embedding model trained with contrastive learning.}
    \label{fig:muq_mulan}
\end{figure}

MuQ-MuLan consists of a music encoder and a text encoder. During training, the music and the corresponding text are fed to the encoders separately, where the text is a description of the music track, as shown in Figure \ref{fig:muq_mulan}.

For music modality inputs, a MuQ model with pre-trained parameters is first used to encode the audio into a latent representation. This representation is then average-pooled over the temporal dimension. Finally, it passes through a linear projection layer to produce a music embedding $e_\mathrm{m}$ with dimension $d$.

For text modality inputs, they are encoded by a pre-trained RoBERTa \cite{liu2019roberta} model. As described in a previous study \cite{vasilakis2024canlisten}, the text encoder is critical to the performance of the two-tower model, so we append a few additional layers of Transformer encoder after RoBERTa. Likewise, the textual latent representations are average-pooled and passed through a linear projection layer to get text embedding $e_\mathrm{t}$ with the same dimension $d$.

We use decoupled contrastive learning (DCL) loss \cite{yeh2022decoupled}. Formally, given the $i$-th (music, text) pair in a mini-batch, for the embeddings $(e_\mathrm{m}^{(i)}, e_\mathrm{t}^{(i)})$, we have

\begin{equation}
   L_{\mathrm{DCL}} = - \mathrm{log} \sum_{(i, j)}\frac{\mathrm{sim}(e_\mathrm{m}^{(i)}, e_\mathrm{t}^{(j)}) / \sigma}{ \sum_{i \neq j} \mathrm{sim}(e_\mathrm{m}^{(i)}, e_\mathrm{t}^{(j)}) / \sigma}
\end{equation}

where $\mathrm{sim}$ denotes dot-product similarity, and $\sigma$ is temperature coefficient.

\section{Experiments Setup}

\subsection{Model Settings}

\subsubsection{MuQ}

We stack 12 layers of Conformer in MuQ, with 310M parameters in total. The number of codebooks for Mel-RVQ is set to $N=8$ and the codebook size is $K=1024$. The input audio is fixed at 30 seconds with a sampling rate of 24K Hz. Audio inputs are 128-dimensional Mel spectrum, which are pooled to 25Hz before feeding to Conformer. The code is implemented with the Fairseq toolkit \cite{ott2019fairseq}.   

\subsubsection{MuQ-MuLan} 

We use the pre-trained MuQ as the initialization for the music encoder. A pre-trained multilingual RoBERTa model 
\texttt{xlm-roberta-base}
is used for the text encoder followed by 8 layers of Transformer, with the embedding dimension $d=512$. MuQ-MuLan has 630M parameters in total, and all parameters stay unfrozen during the contrastive training.

\subsection{Pre-training}
\subsubsection{Dataset}

We initially conduct pre-training on a smaller dataset, Music4all \cite{santana2020music4all}, containing 0.9K hours of open access music data. Subsequently, we scale up to a large-scale in-house dataset with 160K hours of music. 

\subsubsection{Training Details}
Mel-RVQ is trained on the Music4all dataset using 1 GPU. Since the Mel-RVQ is very small (less than 0.3M), the training of Mel-RVQ is extremely fast and it takes less than 1 hour to complete.

 All pre-training experiments are running with 32 A100-40G GPUs. Batch size is 192, and masking probability $p = 0.6$. For the pre-training on the Music4all dataset, we train for only 75K steps. For the experiments on the 160K hours of in-house data, we initially train MuQ for 200K steps. Subsequently, the $l=10$th layer latent is used to train $\mathrm{Mel\text{-}RVQ_{iter}}$, followed by 150K steps of iterative training from scratch to obtain $\mathrm{MuQ_{iter}}$. The entire initial \& iterative pre-training takes about 2 weeks.

\subsection{Contrastive Training}
\subsubsection{Dataset} 
The training data for contrast learning comes from 130K hours of music in-house, each with its corresponding text description. The text descriptions are limited to non-personal information such as genre, mood, instrument, tempo, keywords of the music.

\subsubsection{Training Details}

MuQ-Mulan is trained on 32 V100-32G GPUs with a batch size of 768. Following the practice of MuLan \cite{huang2022mulan}, we switch to taking 10 seconds as the length of the input audio. This significantly reduces memory consumption and enlarges batch size, which is critical for contrastive learning.


\setlength{\tabcolsep}{1mm}
\begin{table*}[h]
    \centering
    \caption{Evaluation of proposed MuQ model in MARBLE benchmark (1/2). $\mathrm{MuQ_{m4a}}$ denotes MuQ model pre-trained on Music4all, a relatively low resource dataset with 900 hours of music. $\mathrm{MuQ_{iter}}$ denotes MuQ model using iterative refinement training. * indicates that the number is brought from the original paper.}
    \begin{tabular}{l|llllllll}
    \toprule[1.5pt]
    \textbf{Dataset} & \textbf{GTZAN} & \textbf{GS} & \multicolumn{2}{c}{\textbf{EMO}}\:\: & \multicolumn{2}{c}{\textbf{VocalSet}}\:\:\:\: & \multicolumn{2}{c}{\textbf{HX}} \\
    \textbf{Task}  & Genre & Key & \multicolumn{2}{c}{Emotion}\:\: & Singer & Tech &  \multicolumn{2}{c}{Structure} \\  
    \hline \rule{0pt}{2.5ex}
    \textbf{Metrics} & Acc$\uparrow$ & Acc\textsuperscript{Refined}$\uparrow$ & R2\textsuperscript{V}$\uparrow$ & R2\textsuperscript{A}$\uparrow$ & Acc$\uparrow$ & Acc$\uparrow$ &  Acc$\uparrow$ & \\ 
    \hline \rule{0pt}{2.3ex}
        MERT & 78.6\textsuperscript{*} & \textbf{65.6}\textsuperscript{*} & 61.2\textsuperscript{*} & 74.7\textsuperscript{*} & 87.1\textsuperscript{*} & 76.9\textsuperscript{*} &   75.7  & \\
       MusicFM & 83.8 & 63.9 & 60.3 & \textbf{76.3} & 92.0 & 78.4 &   76.5  &  \\
       MuQ\textsubscript{m4a} & 84.0\spm{0.3} & 63.7\spm{0.4} & 60.0\spm{0.1} & 76.0\spm{0.1} & 95.2\spm{0.1} & 80.8\spm{0.3} &  76.4\spm{0.1}  & \\
       MuQ & 85.5\spm{0.9} & 63.5\spm{0.7} & 60.9\spm{0.2} & 75.4\spm{0.2} & 95.8\spm{0.1} & \textbf{82.4}\spm{0.3} &  76.5\spm{0.1}  & \\
       MuQ\textsubscript{iter} & \textbf{85.6}\spm{0.7} & 65.0\spm{0.8} & \textbf{62.8}\spm{0.1} & 76.1\spm{0.1} & \textbf{96.2}\spm{0.1} & 81.6\spm{0.1} & \textbf{77.2}\spm{0.1} &  \\
    \bottomrule[1.5pt]
    \end{tabular}
    \label{tab:marble_res1}
\end{table*}

\newcommand{\NA}{\mathrm{NA}}

\setlength{\tabcolsep}{1mm}
\begin{table*}[h]
    \centering
    \caption{Evaluation of proposed MuQ model in MARBLE benchmark (2/2). Avg. denotes the average score calculated across the metrics of all tasks. Pretrain dataset identifies the source and size of the dataset for the pre-training stage of each model.}
    \begin{tabular}{l|llll|c|cc}
    \toprule[1.5pt]
    \textbf{Dataset}  & \multicolumn{2}{c}{\textbf{MTT}\:\:} & \multicolumn{2}{c|}{\textbf{Nsynth}\:\:\:\:} & \multirow{3}{*}{\textbf{\:\:Avg.\:\:}}  & 
 \multicolumn{2}{c}{\multirow{2}{*}{ \textbf{Pretrain dataset}  }}  \\
    \textbf{Task}   & \multicolumn{2}{c}{Tagging}\:\: & Instrument & Pitch &  \\  
    \cline{1-5} \cline{7-8} \rule{0pt}{2.5ex}
    \textbf{Metrics} & ROC$\uparrow$ & AP$\uparrow$
    & Acc$\uparrow$ & Acc$\uparrow$ &   & Source & Size (hours) \\ 
    \hline \rule{0pt}{2.3ex}
        MERT & 91.3\textsuperscript{*} & 40.2\textsuperscript{*}  & 72.6\textsuperscript{*} & \textbf{94.4}\textsuperscript{*} & 74.4 &  - & 160K \\
       MusicFM & 91.3 & 40.0  & 76.2 & 91.1 & 75.4 & MSD & - \\
       MuQ\textsubscript{m4a} &  91.1\spm{0.0} & 39.0\spm{0.1}  & 76.9\spm{0.2} & 91.2\spm{0.1} & 75.8 & Music4all & \textbf{0.9K} \\
       MuQ &  \textbf{91.4}\spm{0.0} & \textbf{40.3}\spm{0.1} & 79.2\spm{0.1} & 92.3\spm{0.2} & 76.7 & In-house & 160K \\
       MuQ\textsubscript{iter}  &  \textbf{91.4}\spm{0.0} & 40.1\spm{0.1} & \textbf{79.7}\spm{0.2} & 91.3\spm{0.1} & \textbf{77.0} & In-house & 160K \\
    \bottomrule[1.5pt]
    \end{tabular}
    \label{tab:marble_res2}
\end{table*}

\subsection{Downstream Tasks}

We evaluate MuQ on the MARBLE benchmark \cite{yuan2023marble} and conduct experiments on a total of \NumOfTask downstream tasks. Note that all datasets appearing in the downstream tasks are \textbf{not} included in the pre-training data of MuQ.

Due to page limitations, we only provide a brief overview of the downstream tasks here. We encourage readers to refer to the Appendix  \uppercase\expandafter{\romannumeral1} for details on these tasks.

\begin{itemize}
    \item \textbf{Genre classification} aims to determine the most suitable genre for a song. We use the GTZAN dataset \cite{Tzanetakis_Cook_2002} and report the accuracy scores.
    \item \textbf{Key detection} predicts the tonal scale and dominant pitch level. We use the Giantsteps dataset \cite{Knees_2015_GS}, and a refined accuracy metric \cite{raffel2014mir_eval}.
    \item \textbf{Emotional analysis} is taken on the Emomusic dataset \cite{Soleymani_Caro_Schmidt_Sha_Yang_2013}.  The metrics are determination coefficients for valence ($\mathrm{R2^V}$) and arousal ($\mathrm{R2^A}$) score.
    \item \textbf{Singer identification} on the VocalSet \cite{Wilkins_2018_vocalset}
dataset. The evaluation metric is accuracy.
    \item \textbf{Vocal technique detection} on the VocalSet \cite{Wilkins_2018_vocalset}. The evaluation metric is accuracy.
    \item \textbf{Music tagging} involves multi-label classification of 50 tags on the MagnaTagATune \cite{law2009mtt} dataset. The metrics are ROC-AUC and average precision (AP).
    \item \textbf{Music structure analysis} predicts the functional
label for music segments on the Harmonix dataset \cite{nieto2019harmonix}. The evaluation metric is accuracy.
    \item \textbf{Instrument classification} on the Nsynth dataset \cite{Engel_2017_NSynth}. The evaluation metric is accuracy.
    \item \textbf{Pitch classification} of the 128-pitch on the Nsynth dataset \cite{Engel_2017_NSynth}. The metric is accuracy.
\end{itemize}

In evaluating the pre-trained MuQ, we extract features from each layer and use a linear probe to perform downstream tasks. For the MuQ-MuLan experiments, we use a zero-shot approach for music tagging. This involves directly calculating the similarity between music and text embeddings to determine the prediction probability for each tag.

\section{Results and Analysis}

\subsection{Performance of Pre-trained MuQ}

We compare MuQ with two popular models, MERT and MusicFM. For MERT, we use \texttt{MERT-v1-330M}
version \cite{li2023mert}, the most superior of its releases on HuggingFace. For MusicFM, we use the checkpoint pre-trained on the MSD dataset \cite{bertin2011msd}, which is publicly available by \cite{won2023musicfm}. All models are of around 330M parameters.

Table \ref{tab:marble_res1} and Table \ref{tab:marble_res2} present the results for three versions of MuQ. Considering the experiments are very extensive and time-consuming, we conduct 5 runs for MuQ models and report the mean scores and standard deviations for all tasks.

It is shown that for MuQ\textsubscript{m4a}, pre-trained on the 0.9K-hour Music4All dataset, already outperforms MERT and MusicFM. Training MuQ on a larger dataset further amplifies this advantage. Iterative training slightly improves the effectiveness of MuQ, resulting in MuQ\textsubscript{iter} with an average score of 77.0, which is our best-performing model. 

\begin{figure}[h]
    \centering
    \includegraphics[width=1\linewidth]{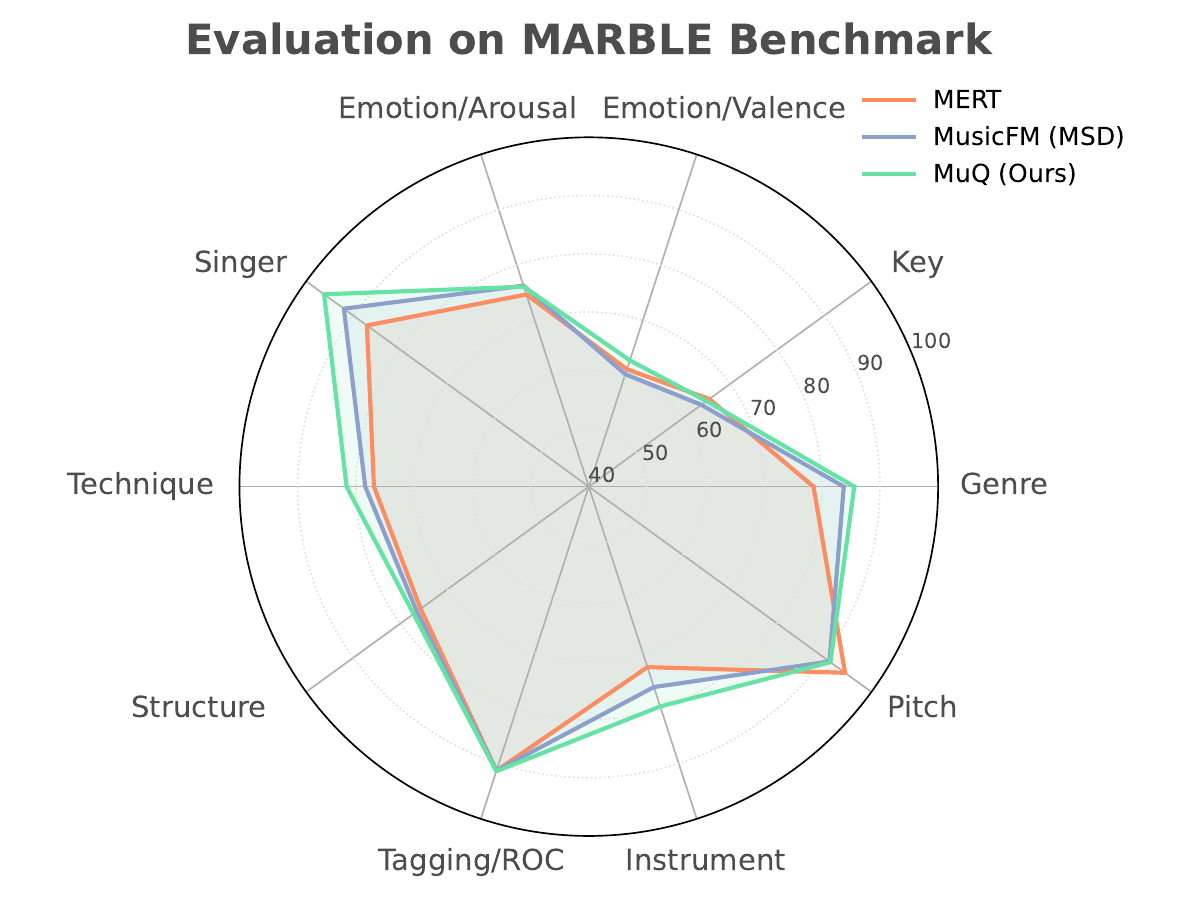}
    \caption{Visualization of radar plot on MARBLE benchmark.}
    \label{fig:radar}
\end{figure}

MuQ outperforms previous models on many downstream tasks, especially in genre classification, singer identification, vocal technique detection, music structure analysis and instrument classification. We visualize the result of MuQ in Figure \ref{fig:radar}.

\subsection{Ablation Study on Mel-RVQ}

\setlength{\tabcolsep}{1mm}
\begin{table*}[h]
    \centering
    \caption{Ablation study of different components in Mel-RVQ. All models are pre-trained on the Music4all dataset.}
    \begin{tabular}{c|c|c|cccccc|c}
    \toprule[1.5pt]
    & \multicolumn{2}{c|}{\textbf{Components}} & \textbf{GTZAN} & \textbf{GS} & \multicolumn{2}{c}{\textbf{EMO}} & \multicolumn{2}{c|}{\textbf{VocalSet}} & \\ \cline{2-3}
     &&& Genre & Key & \multicolumn{2}{c}{Emotion} & Singer & Tech &  \\ 
    \multirow{-3}{*}{\begin{tabular}[c]{@{}c@{}}\textbf{Model} \\ \textbf{codebase}\end{tabular}} &
    \multirow{-2}{*}{\textbf{VQ type}}&\multirow{-2}{*}{\textbf{ Residual}}  &
    Acc$\uparrow$ & Acc\textsuperscript{Refined}$\uparrow$ & 
    R2\textsuperscript{V}$\uparrow$ & R2\textsuperscript{A}$\uparrow$ & 
    Acc$\uparrow$ & Acc$\uparrow$ & \multirow{-3}{*}{\textbf{Average}}   \\ 
    \hline \rule{0pt}{2.3ex}
      & Random & \usym{2717} & 77.9 & 62.8 & 56.4 & 71.8 & 95.5 & 76.3 & 73.5  \\
      \multirow{-2}{*}{MusicFM} & Trained & \usym{2717}  & 80.0 & \textbf{65.8} & 56.3 & 74.6 & 95.3 & 78.5 & 75.1  \\
      \hline \rule{0pt}{2.3ex}
      & Random & \usym{1F5F8}  & 77.6 &62.1 & 56.5 & 74.1 & \textbf{95.7} & 76.5 & 73.8  \\
    \multirow{-2}{*}{MuQ} & Trained & \usym{1F5F8}  & \textbf{84.0} & 63.7 & \textbf{60.0} & \textbf{76.0} & 95.2 & \textbf{80.8} & \textbf{76.6}  \\
    \bottomrule[1.5pt]
    \end{tabular}
    \label{tab:comp_abl}
\end{table*}

We carry out an ablation study for Mel-RVQ in Table \ref{tab:comp_abl} to demonstrate the effect of its key components. 
In the ablation experiments, all models are pre-trained 75K steps on the Music4all with 310M parameters, and tested on 5 tasks. 

The VQ type in Table \ref{tab:comp_abl} indicates whether the VQ is randomly initialized or trained on music data, and Residual indicates whether residual multiple codebooks are used. As shown, it's crucial to train the VQ rather than leaving it random, as this boosts the performance of MusicFM and MuQ by 1.6 and 2.8, respectively. Also, the residual structure of Mel-RVQ is beneficial and allows MuQ to gain a significant advantage over MusicFM.

\subsection{Ablation Study on Music SSL Target}

\setlength{\tabcolsep}{1mm}
\begin{table*}[h]
    \centering
    \caption{Ablation study of target strategy. All models are pre-trained on Music4all dataset.}
    \begin{tabular}{l|c|cccccc|c}
    \toprule[1.5pt]
     & & \textbf{GTZAN} & \textbf{GS} & \multicolumn{2}{c}{\textbf{EMO}} & \multicolumn{2}{c|}{\textbf{VocalSet}} & \\
      & & Genre & Key & \multicolumn{2}{c}{Emotion} & Singer & Tech &  \\  
    \multirow{-3}{*}{\textbf{Tokenizer}} & \multirow{-3}{*}{\begin{tabular}[c]{@{}c@{}}\textbf{Num. of} \\ \textbf{codebooks}\end{tabular}} & Acc$\uparrow$ & Acc\textsuperscript{Refined}$\uparrow$ & R2\textsuperscript{V}$\uparrow$ & R2\textsuperscript{A}$\uparrow$ & Acc$\uparrow$ & Acc$\uparrow$ & \multirow{-3}{*}{\textbf{Average}}   \\ 
    \hline \rule{0pt}{2.3ex}
       Mel-RVQ & 1 & 82.8 & 45.9 & 55.3 & 74.6 & 93.7  & 75.4 & 71.3  \\
       \;Mel-RVQ & 4 & 78.3 &  65.1  & 59.0  &  73.7 & 95.2  & 79.5 & 75.1 \\
       \;Mel-RVQ & 8 & \textbf{83.4}  & \textbf{65.8} & \textbf{59.7} & \textbf{76.2} & \textbf{95.7} & \textbf{80.1} & \textbf{76.8}\\
       \;Mel-RVQ (deeper) & 8 & 81.7 & 64.4  & 55.9  &  72.3 & 95.5  & 80.4 & 75.0  \\
       \;Encodec & 8 & 82.4  &  65.3 &  57.1  & 74.1  & 95.2  & 77.9 & 75.3   \\
    \bottomrule[1.5pt]
    \end{tabular}
    \label{tab:target_abl}
\end{table*}

We ablate the number of target tokens and tokenization SSL training, this involves some hyperparameters of Mel-RVQ, as well as a comparison with Encodec target.

As seen in Table \ref{tab:target_abl}, performance gradually improves as the number of codebooks increases. We believe this is because more codebooks allow the model to capture fine-grained audio features more effectively.  This also implies the advantage of multi-target learning for SSL pre-training.

We also explore the depth of the Mel-R\textbf{}VQ encoder and decoder by scaling up the linear projection to a 3-layer fully connected network, referred to as Mel-RVQ (deeper)  in Table \ref{tab:target_abl}. However, the results suggest that increasing the depth of the Mel-RVQ network actually degrades performance. This indicates that a single layer of linear projection is sufficient for Mel-RVQ to achieve optimal performance. 

Finally, we compare Mel-RVQ with the Encodec target.  It can be seen in Table \ref{tab:target_abl} that Mel-RVQ, despite being a very lightweight structure, outperforms the Encodec tokenization target. We attribute this to the fact that the token produced by Encodec is a high abstraction of the music audio, which leads to it being more difficult for the model to learn.

\subsection{Performance of MuQ-MuLan}

In Table \ref{tab:mulan} , we compare MuQ-MuLan with LAION-CLAP \cite{laionclap2023}, Microsoft-CLAP 2023\cite{CLAP2023} and MuLan \cite{huang2022mulan}. We use the open-source \texttt{larger\_clap\_music\_and\_speech} version for LAION-CLAP, which is the best-performing checkpoint among its releases in our experiments. We also introduce two variants that employ MERT or MusicFM as the music encoder and refer to them as MERT-MuLan and MusicFM-MuLan.

Following previous study \cite{vasilakis2024canlisten}, we choose the proven superior MusCALL prompt: ``A \textless label\textgreater \: track" for all models to test in our experiments.

\setlength{\tabcolsep}{1mm}
\begin{table}[h]
    \centering
    \caption{Experiment on the MagnaTagATune zero-shot music tagging task. Since MuLan is a closed source, the results of MuLan are taken from the paper \cite{huang2022mulan}.}
    \begin{tabular}{l|c|cc}
    \toprule[1.5pt]
         \multirow{2}{*}{\textbf{Model}} & \multirow{2}{*}{\begin{tabular}[c]{@{}c@{}}\textbf{Data} \\ \textbf{(hours)}\end{tabular}}  &   \multicolumn{2}{c}{  \textbf{Metrics}}   \\
         & &  ROC-AUC$\uparrow$ & PR-AUC$\uparrow$   \\
    \hline 
        LAION-CLAP & 4K & 73.9 & 25.0 \\
        Microsoft-CLAP 2023 & - & 75.9 & 28.9 \\
        MuLan & 370K & \:\;78.2\textsuperscript{*} & -   \\
        \hline
        MERT-MuLan & 130K & 75.5 & 24.6 \\
        MusicFM-MuLan & 130K & 76.2 & 27.1 \\
        MuQ-MuLan & 130K & \textbf{79.3} & \textbf{29.3}   \\
    \bottomrule[1.5pt]
    \end{tabular}
    \label{tab:mulan}
\end{table}

MuQ-MuLan outperforms previous models in both ROC-AUC and PR-AUC and reaches the SOTA in the zero-shot setting. 
Impressively, MuQ-MuLan still has a slight advantage over MuLan despite the gap in data amounts. 

Although both use the same contrastive learning setup, MuQ-MuLan outperforms MERT-MuLan. Furthermore, even though MuQ and MusicFM have identical model structures,  MuQ-MuLan still performs better. This suggests that MuQ provides a better initialization for contrastive training. 

\begin{figure}
    \centering
    \includegraphics[width=1\linewidth]{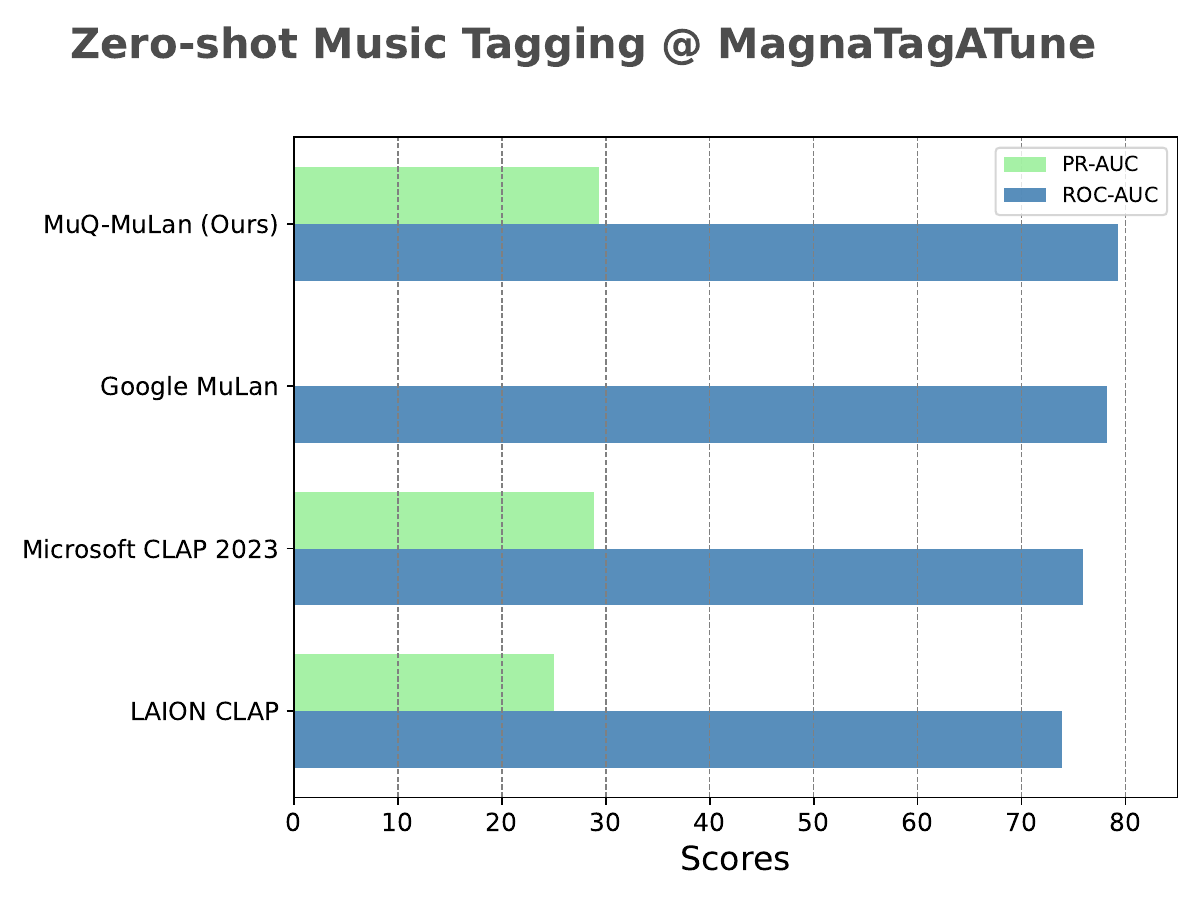}
    \caption{Visualization on zero-shot music tagging task in MagnaTagATune dataset.}
    \label{fig:tagging}
\end{figure}

We visualize the performance of MuQ-MuLan in Figure \ref{fig:tagging}, and plot the comparision with LAION-CLAP \cite{laionclap2023}, Microsoft-CLAP 2023 \cite{CLAP2023} and MuLan \cite{huang2022mulan}.

\subsection{Layer-wise Analysis}

\begin{figure}[h]
    \centering
    \includegraphics[width=1\linewidth]{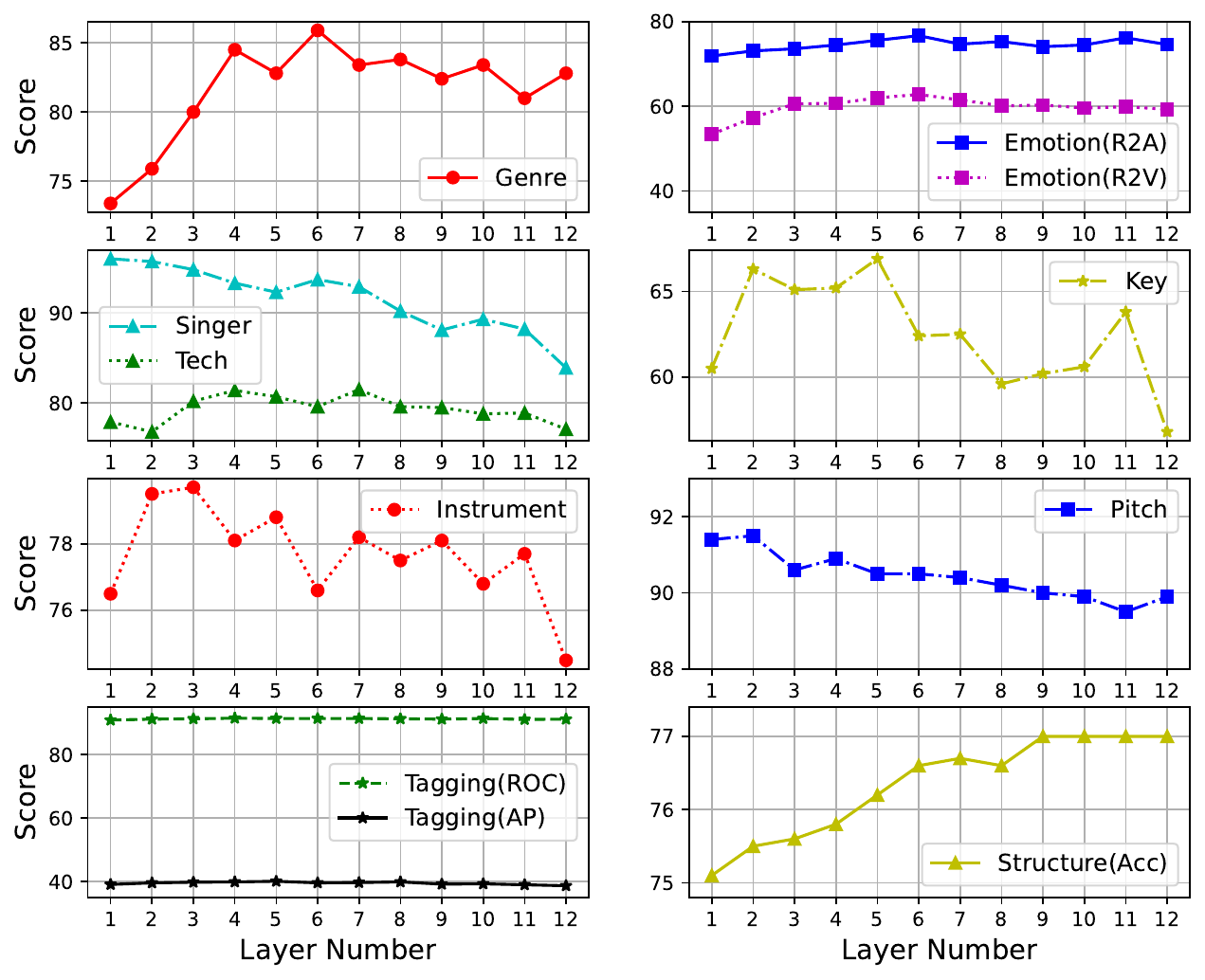}
    \caption{Layer-wise results of MuQ in MARBLE benchmark.}
    \label{fig:task_layer}
\end{figure}

In Figure \ref{fig:task_layer}, we visualize the results of each layer's features in various downstream tasks during one run of the MuQ model. The analysis can be summarized as follows:

\begin{itemize}
    \item \textbf{Semantic tasks}, such as genre classification and structure analysis, generally achieve the best results at the higher layers. They focus more on semantic information such as the direction of the music and the overall style.
    \item \textbf{Acoustic tasks}, such as singer identification, instrument classification, pitch classification, and key detection, perform best at the lower layers. This implies that they are more closely associated with low-level characteristics such as the original audio features.
    \item \textbf{Comprehensive tasks}, such as music tagging, emotional analysis, and vocal tech classification, have a relatively even distribution across all layers. This suggests that they rely on various aspects.
\end{itemize}


\section{Discussion}

\paragraph{Additional explanation of Mel-RVQ}
Although in this paper, we only discuss the case of Mel-RVQ with up to 8 codebooks, this does not mean that more codebooks are meaningless. Exploring Mel-RVQ with more codebooks may help understand the reasons why Mel-RVQ works and the limits of this advantage.
The selection of layer 10 in iterative training is based on empirical intuition rather than rigorous experimental comparison. Due to limited computational resources, we did not explore other layers. We believe that iterating over other layers could potentially lead to better results, and this is a worthwhile topic of future research.

\paragraph{Potential role of virtual classes in SSL} 

The concept of virtual classes, introduced in \cite{chen2018virtual}, suggests that adding dummy classes to classification tasks can improve performance. We note that in fact the random projection quantizer in BEST-RQ, while having low codebook utilization, those codes that are never hit may in fact provide a role similar to that of virtual classes. In addition, we also observe in our early experiments with Mel-RVQ that virtual classes have a potential effect on the SSL training. Although our approach does not involve virtual classes,  we believe their role in SSL merits further investigation.

\paragraph{Limitation and future work} 

Mel-RVQ is required to be pre-trained on music data before being used for SSL targeting. While we emphasize that the training process for Mel-RVQ is very fast (less than 1 hour), this is still not a completely out-of-the-box approach like BEST-RQ. In future work, we will also explore different model sizes (e.g., 90M $base$ size) for the requirements of different physical environments and computational conditions.

\section{Conclusion}
In this study, we introduce MuQ, a novel self-supervised learning model for music representation and understanding. Building on this, we also present MuQ-MuLan, a joint music-text embedding model. MuQ uses Mel-RVQ as a quantization target, applying a linear projection RVQ structure to the Mel spectrum, which provides a simple yet powerful SSL target. Our model shows significant improvements over previous models like MERT and MusicFM, as demonstrated by its performance on the MARBLE benchmark. Additionally, experiments on the zero-shot music tagging task reveal that MuQ-MuLan achieves state-of-the-art music-text modal alignment. Our model code and checkpoints will be made open-source to support future research.

\section{Ethic Discussion}

We recognize the potential ethical concerns associated with MuQ. Throughout the data handling process, we’ve made it a priority to follow ethical guidelines, ensuring that personal information is protected and that the model remains free from biases. We also guarantee that no audio will be distributed or shared. We believe our work can positively impact the community by advancing automated music analysis, and MuQ can serve as a valuable tool for musicians and creators.

\bibliography{aaai25}

\newpage

\section{Appendix \uppercase\expandafter{\romannumeral1}: Evaluation Details}

\subsection{Downstream Tasks}

We present here a detailed description of 9 downstream tasks involved in the evaluation, with their respective datasets and evaluation metrics. Most of the downstream tasks are based on the awesome MARBLE-benchmark \cite{yuan2023marble} implementation.

\subsubsection{Genre classification} 
Music Genre classification is an important tasks of Music Information Retrieval (MIR). It aims to determine the most suitable genre for each song in dataset. We use the GTZAN dataset \cite{Tzanetakis_Cook_2002} for this task. 

The GTZAN dataset has a collection of 10 genres with 100 audio files each, all having a length of 30 seconds. We set windows of 10-seconds as inputs to get the features respectively, then combine the results to get the features used for the downstream task. We utilize the standard ``fail-filtered" split \cite{Kereliuk_Sturm_Larsen_2015} for training and evaluating, then we provide the accuracy scores on testing set to show the final result.

\subsubsection{Key detection} 
Key detection estimates the tonal scale and dominant pitch level of each given song. Specifically, the labels of the task contains major and minor scales for all pitch classes, that is, 24 classes in total.  

We use the Giantsteps dataset \cite{Knees_2015_GS}, which has a collection of annotations for 604 2-min audio previews of songs, as the test set, and a commonly-used subset of the Giantsteps-MTG-keys dataset \cite{Korzeniowski_Widmer_2017}, for training and validation. We used the split method that was described in \cite{Castellon_Donahue_Liang_2021}. Finally, we used the refined accuracy with error tolerance score \cite{raffel2014mir_eval} following Table \ref{tab:r_acc} to evaluate the performance of models.

\begin{table}[h]
    \centering
    \caption{Calculation method of Refined Acc score }
    \begin{tabular}{l|c}
    \toprule[1.5pt]
         \textbf{Relationship } & \textbf{Score}   \\
    \hline 
        Same key and mode & 1.0 \\
        Estimated key is a perfect fifth above reference &  0.5 \\
        Relative major/minor(same key signature)  &  0.3 \\
        Parallel major/minor(same key)  &  0.2 \\
        Other  &  0.0  \\
        
    \bottomrule[1.5pt]
    
    \end{tabular}
    \label{tab:r_acc}
\end{table}

\subsubsection{Emotional analysis}
The Music Emotion Recognition aims to get the Emotion score of songs. We use the Emomusic dataset \cite{Soleymani_Caro_Schmidt_Sha_Yang_2013}, which contains music clips of 45 seconds. The dataset has 1000 clips in total. After removing the duplicates, the size of the dataset is reduced to 744 songs. Each clip is labeled with two scores: valence score indicates positive and negative emotional responses, and arousal score indicates emotional intensity. The dataset split is same as \cite{Castellon_Donahue_Liang_2021}.We divide the whole signal into 5s windows as the inputs of SSL models to get the features, then combine the features as the inputs of a simple regression models.

At testing stage, we can get multiple predictions. Each prediction of a given song is a pair of scores of valence score and arousal score, which is same as the ground truth. We calculate the determination coefficient (R2) between the model regression results and human annotations of arousal (R2A) and valence (R2V) \cite{Soleymani_Caro_Schmidt_Sha_Yang_2013} as metrics for the test results.

\subsubsection{Singer identification}

Singer identification is to recognize the corresponding singer based on the given singing audio.

We use the Vocalset dataset \cite{Wilkins_2018_vocalset}, which contains 20 different singers (11 male and 9 female) who perform 17 different singing techniques in various contexts. We randomly split the dataset into traning,validation and testing sets based on a ratio of 12:8:5, all of which contain the full 20 singers. The evaluation metric is accuracy. 

\subsubsection{Vocal technique detection}
Vocal technique detection is to identify the techiques that the singer used in a given audio segement. We use the VocalSet dataset for this task. As the audio clips are divided into 3 seconds, the task only requires a judgement on the type of technique and not on the start and end of the technique. We used the same 10 different singing techniques as in \cite{Yamamoto_Nam_Terasawa} as a subset and used the same 15 singers as the training and validation sets and 5 singers as the test set. Due to the lack of a standardized division between training and validation sets, we opted for a selection of 9 singers as the training set and 6 singers as the validation set. All segments originating from the same recording are assigned to the same portion of the split. For evaluation purposes, we employ accuracy as the metric in this classification task.

\subsubsection{Music tagging}
Music tagging assigns a set of fixed tags for a given song. Tag contains genre, instrumentation, mood and so on. We use the MagnaTagATune (MTT) \cite{Law_West_Mandel_Bay_Downie_2009} dataset. MTT dataset contains 25,863 music clips. Each clip is a 29-seconds-long excerpt belonging to one of the 5223 songs, 445 albums and 230 artists.  The clips span a broad range of genres like Classical, Rock, Pop, Jazz, Blues and more. Each audio clip is labeled by human with a vector of binary annotations of 188 tags. The annotations include tags like `singer’, `no singer’, `violin’, `drums’, `classical’, `jazz’. The top 50 most popular tags are typically used for evaluation to ensure that there is enough training data for each tag. 

For both datasets, we limit the tag vocabulary according to official instructions. We use all clips in MTT. The metrics are the macro-average of ROC-AUCs, a multiclass AUC as the average of the area under multiple binary ROC curves, and the average precision (AP) / PR-AUC, the weighted mean of precisions at each threshold, among all top-50 tags.

\subsubsection{Music structure analysis}
aims to split a music signal into non-overlapping sections and predict the functional label for each segment, including `intro',`verse',`chorus' and so on. We use the Harmonix Set \cite{nieto2019harmonix} for the experiment. This is a frame-wise classification task. We first construct a probability curve based on the given data annotations, where each frame contains a boundary score and a function score.

To standardize the labels, we mapped the labels in the data following the setting in \cite{Wang_Hung_Smith_2022}. Then, to ensure smooth training of the model, we create smoothed function and boundary activation curves. We smooth the transitions of the function activation curves using a Hanning window: a 1-second ramp from 0 to 1 prior to the onset, and a 1-second ramp down after the offset, as in \cite{Wang_Smith_Chen_Song_Wang_2021}. For boundary curves, we also set a duration of 0.6 seconds for each boundary using a Hanning window.

We use a simple LSTM model to predict frame-wise probabities of functional and boundary curves. Frame-wise accuracy for functional labels is used as metric.

\subsubsection{Instrument classification}

Instrument classification identify which instruments are used in a given music segment. We use the NSynth dataset \cite{Engel_2017_NSynth} and MTG-instrument dataset.The NSynth Dataset is an audio dataset containing ~300k musical notes, each with a unique pitch, timbre, and envelope. Each note is annotated with three additional pieces of information based on a combination of human evaluation and heuristic algorithms: Source, Family, and Qualities. Each segment may contain multiple instruments. We evaluate the model using ROC and AP scores.

\subsubsection{Pitch classification}

Pitch classification aims to detect the certain pitch of a given short audio clip in 128 pitch categories. We use the NSynth dataset \cite{Engel_2017_NSynth} for this task, then we use accuracy score to evaluate the performance.

\end{document}